\documentstyle[11pt]{article}
\def\ZZ{{\mathchoice {\hbox{$\sf\textstyle Z\kern-0.4em Z$}}
{\hbox{$\sf\textstyle Z\kern-0.4em Z$}}
{\hbox{$\sf\scriptstyle Z\kern-0.3em Z$}}
{\hbox{$\sf\scriptscriptstyle Z\kern-0.2em Z$}}}}

\newcommand{\be}{\begin{equation}}
\newcommand{\ee}{\end{equation}}

\def\fun#1#2{\lower3.6pt\vbox{\baselineskip0pt\lineskip.9pt
\ialign{$\mathsurround=0pt#1\hfil##\hfil$\crcr#2\crcr\sim\crcr}}}

\begin{document}
\thispagestyle{empty}
\noindent\hspace*{\fill}  FAU-TP3-98/15  \\
\noindent\hspace*{\fill}  hep-th/yymmdd \\
\noindent\hspace*{\fill}  15 August, 1998 %\today   \\

\begin{center}\begin{Large}\begin{bf}
Confining Properties of the
Homogeneous Self-Dual Field \\
and the Effective Potential in SU(2) Yang-Mills Theory.   \\

\end{bf}\end{Large}\vspace{.75cm}
 \vspace{0.5cm}

Garii V. Efimov\footnote{efimovg@thsun1.jinr.r}, \\
Bogoliubov Laboratory of Theoretical Physics, \\
Joint Institute for Nuclear Research, 141980 Dubna,
Russia, \\
Alex C. Kalloniatis %
\footnote{ack@theorie3.physik.uni-erlangen.de}
\\
Institut f\"ur Theoretische Physik III
Universit\"at Erlangen - N\"urnberg \\
Staudtstra{\ss}e 7
D-91058 Erlangen, Germany
\\

and Sergey N. Nedelko\footnote{nedelko@thsun1.jinr.ru}, \\
Bogoliubov Laboratory of Theoretical Physics, \\
Joint Institute for Nuclear Research, 141980 Dubna,
Russia \\

and \\

Institut f\"ur Theoretische Physik III
Universit\"at Erlangen - N\"urnberg \\
Staudtstra{\ss}e 7
D-91058 Erlangen, Germany
\end{center}
\vspace{1cm}\baselineskip=35pt

\date{\today}
\begin{abstract} \noindent
We examine in non-Abelian gauge theory
the heavy quark limit in the presence
of the (anti-)self-dual homogeneous background field and see that
a confining potential emerges, consistent with the Wilson criterion,
although the potential is quadratic and not linear
in the quark separation. This
builds upon the well-known feature that propagators in
such a background field are entire functions.
The way in which deconfinement can occur at finite
temperature is then studied in the static temporal gauge
by calculation of the effective potential at high temperature.
Finally we discuss the problems to be surmounted
in setting up the calculation of the
effective potential nonperturbatively on the lattice.
\end{abstract}
\newpage\baselineskip=18pt

\section{Introduction}
Over the years, various characterisations
have been proposed for 'confinement', the
property that coloured degrees
of freedom are undetectable at present-day collider
energies. Certainly, the Wilson criterion
that static colour sources cannot be separated arbitrarily
far apart \cite{Wil74} has lead to many insights both in lattice
simulations
and in analytical calculations. In particular, based on the
Wilson criterion lattice simulations have established
a confinement-deconfinement phase transition at finite
temperature.
There are however alternate characterisations for confinement
which may be more directly relevant for dynamical quarks and
gluons,
and which are based on the analytic properties of the
nonperturbative quark or gluon propagators.
In this paper, we shall focus on the suggestion that
the absence of poles in the complex energy plane of
field propagators is consistent with
confinement of quarks and gluons, in other words that
propagators are entire functions. That this can be correctly
described as `confinement' is easy to see: absence of poles means
that no  coloured degrees of freedom can appear
in physical asymptotic states. This characterisation
of confinement is not necessarily in conflict with the Wilson
criterion. Indeed, one of our aims will be to show
that, in the static quark limit, entire quark propagators lead
to
the Wilson criterion.

A quite simple mechanism for rendering
quark and gluon propagators entire in the complex energy plane
is to apply a homogeneous background gluon field which satisfies
the key property that it be either self-dual or anti-self-dual.
Such a background gauge field is characterised by
\begin{eqnarray}
B^a_\mu(x) \tau^a & \equiv &
\frac{1}{2} n^a \tau^a B_{\mu \nu} x_\nu
\label{field}  \\
{\tilde B}_{\mu \nu} & = & \frac{1}{2} \epsilon_{\mu \nu \rho
\lambda}
B_{\rho \lambda} = \pm B_{\mu \nu}
\label{duality} \\
B_{\mu \nu} B_{\mu \rho} & = & B^2 \delta_{\nu \rho}, \ B =
{\rm{const}},
\nonumber\\
B_{ij} & = & - \epsilon_{ijk} B_k, \ B_{j4} = \pm B_j
\label{cptduality}
\ .
\end{eqnarray}
The positive and negative sign in
Eqs.(\ref{duality},\ref{cptduality})
correspond, respectively, to the self-dual and anti-self-dual
cases.
The colour vector $n^a$ points in some fixed direction
which can be chosen such that $n^a \tau^a$
is diagonal; $n^a$ picks out the Cartan subalgebra
of the colour group.
Various properties of this field in SU(2) gauge theory
were investigated originally in \cite{Leu81,Min81}. For example, 
in contradistinction to the chromomagnetic background field
\cite{Sav77} the self-dual background is stable.
Moreover, it was observed
that this field leads to entire functions for the charged scalar
field propagator. In the sense described above then, this
field can provide for confinement of quarks and gluons.
Diagonal components of the gluon field (as SU(N) algebra
elements)
are not confined at least at the level of
the lowest order propagator in the background field. A self-dual
homogeneous
field is at least then a possible source for confinement in QCD
if it can be shown that such a field is a dominant configuration
in the QCD functional integral.

This verification can come from a computation of the
effective potential for the candidate background field and the
demonstration that the potential has a minimum at a nonzero
value for the background field. The effective potential was
calculated to one-loop in \cite{Leu81,Eliz}.
These results however were inconclusive in the sense that
the quantum corrections to the potential
were as large as the zeroeth order classical term.
To our knowledge, despite several attempts to study the
effective potential for the Savvidy chromomagnetic background
\cite{Sav77}
on the lattice \cite{CeC97,CeC91,AMBZ89,LeP95,TrW93},
an analogous nonperturbative computation
for the self-dual homogeneous background has not been attempted.
Nonetheless, with the assumption that the effective potential
for a self-dual homogeneous background field has a nontrivial
minimum and using just those quark and gluon propagators
which exhibit confinement in the sense of entire functions,
some successful phenomenological investigations for SU(3)
have been carried out \cite{EfN98,selfdualphen}. Quantitatively,
experimental data for the spectrum of light, heavy-light and
heavy quarkonium systems can be reproduced to within
ten percent in this effective description.

In this work we concentrate on the problem of confinement and
the effective potential for the SU(2) gauge theory.
Our goal is first to describe
the confining properties of the self-dual background field
in the more familiar terms of the Wilson picture \cite{Wil74}.
Secondly, we seek to show that, even if we cannot prove
the existence of a nontrivial minimum in the effective
potential for this background field at zero temperature and
strong coupling,
nonetheless deconfinement at {\it high temperature} can occur.
Namely, we will show that at high temperature
the effective potential for a self-dual background field acquires
a minimum at zero field value.

In the first instance we
illustrate the confining properties of the self-dual homogeneous
background
by studying the problem of heavy  particle and anti-particle
in this background field.
We thus examine the nonrelativistic limit. We indeed find
that a confining potential for static charges emerges:
the stationary trajectories of particle and anti-particle
in the background field (\ref{field})
separated by distance $|\vec X|$
and held apart for time $T$ are suppressed by a factor
$$\exp\left(-iT\frac{B^2}{32\mu}\vec X^2\right),$$
where $\mu$ is a reduced mass of two-particle system.
This result differs from that seen in most lattice simulations
because of the different long-range properties of the field
considered here as compared to those normally implemented in lattice
gauge theory. The oscillator binding potential
arises here effectively due to an interaction of the charges
with the background field, but not by virtue of quantum gluon
exchange between these charges.
The self-duality and homogeneity of the background
field is of crucial importance. The oscillator
nonrelativistic potential is not inconsistent
with the phenomenology of Regge trajectories in the hadronic
spectrum since the latter is a feature of light quark systems.
In the approach
to the relativistic bound state problem
of \cite{selfdualphen}
based on the bosonisation
of the one-gluon exchange interaction between quark currents in the
presence of the vacuum field~(\ref{field})
it is seen that the property
that quark and gluon propagators be entire
precisely gives rise to Regge behaviour in light-quark systems.

It is appropriate to mention here the evident fact that
a vacuum field such as Eq.(\ref{field}) would lead to breaking
of a range of symmetries such as CP, colour and O(3).
A satisfactory restoration of these symmetries at
the hadronic scale assumes the inclusion of
domain structures in the vacuum. In
a given domain the vacuum field has a specific direction
and is either self-dual or anti-self-dual, but
this is uncorrelated with the specific realisation
of Eq.(\ref{field}) in another domain.
% in which the direction of the vacuum
%field can be different and arbitrary and in which the self-dual
configuration
%can flip to the anti-self-dual one.
The idea of domains in the QCD vacuum was discussed in
application to various homogeneous fields~\cite{Leu81,Amb80,AmS90}.
In the effective meson Lagrangian of \cite{selfdualphen}
this idea was realized as the prescription that
different quark loops (namely, those separated by the meson lines)
in a diagram must be averaged over different configurations of the vacuum
field~(\ref{field}) independently of each other.
In the present paper we do not consider this problem,
and only wish to note that the above formula for the contribution of
stationary
trajectories does not depend on directions and is the
same for both self-dual and anti-self-dual homogeneous fields.

In the second instance, though we cannot compute the
effective potential nonperturbatively we nonetheless seek to
show that at high temperature, where asymptotic freedom
should set in, the effective potential does actually
acquire a minimum at zero external field consistent with
deconfinement. This is not just a trivial consequence of
perturbation theory. Lattice simulations have confirmed the
picture that high temperature Yang-Mills theory, though
deconfined, shows significant signals of nonperturbative
structure \cite{Karsch}.
In order to account for some of these properties we have used
the recent developments in temporal/axial type gauges
at finite extension or temperature by Lenz {\it et al}
\cite{LT98,EKLT98}.
Here a complete gauge fixing of Yang-Mills theory was formulated,
accompanied by an integration out of certain zero mode
fields which themselves are related intimately to the
Polyakov loop order parameter \cite{Sve86}
for the confinement-deconfinement
phase transition in pure Yang-Mills theory. The integration
out of these variables generates for off-diagonal gluon fields
a temperature dependent mass $M(T)$ which diverges with
increasing temperature, $T$. In \cite{LT98} it was checked that, despite
the
gluon mass, renormalisation at the one-loop order was standard,
leading to the correct one-loop beta function for
SU(2) consistent with gauge invariance. Moreover, this mass was
shown to be related to the string constant in a linearly
confining potential. Though the actual mechanism for confinement
in our study is quite independent of that in \cite{LT98},
this gluon mass generation is of crucial importance for us. It
defines a scale $M=M(T)$
in the running coupling constant $g_R(M)$ so that at high
temperature the coupling is small. We are thus able to perform a controlled
calculation and find that at high temperature the effective potential takes
the form,
$$ U_{\rm eff}(B^2) = {{B^2} \over {g^2_R(M)}} +
{{29}\over{525\pi^2}} {{B^4} \over {M^4(T)}}
+ {\cal O}(B^6/M^8(T)) + {\cal O}(g^2_R(M)) $$
which has a minimum at zero field $B=0$. If non-zero $B$ can
generate confinement at zero and low temperatures, then our
result shows that deconfinement at high temperature can occur.

In the following section we demonstrate that the self-dual
homogeneous field provides simultaneously for the Wilson
confinement criterion and the property that
propagators of off-diagonal (charged) fields in a
self-dual homogeneous gauge field
are entire functions. Following that we consider the high
temperature
limit in the effective potential. The paper concludes with a
summary of results and a discussion of the problem of computing
the effective potential on the lattice.
Much of the detail of explicit calculations is relegated to four
Appendices.

\section{Self-dual Homogeneous Field and the Wilson Criterion }
To illustrate the relationship between confinement and the
property that Green's functions in an (anti-)self-dual
background field are entire functions it suffices to consider
a simple charged scalar field of mass $m$ coupled to the
background gauge field $B_\mu=B_{\mu\nu}x_\nu$ defined by
Eqs.(\ref{field}-\ref{cptduality}).
The relationship between this and the original Yang-Mills theory
can be understood as follows: by assumption, the effective potential for
the configuration Eqs.(\ref{field}-\ref{cptduality}) exhibits
a minimum at $B^2 \neq 0$ which itself is proportional to
the fundamental scale of the theory, $\Lambda_{\rm{YM}}$.
By shifting the fields, we study the coupling of small fluctuations
to this non-vanishing background. Thus the $\phi$-fields are
those components of the gluon field which couple in the leading
order to the background. We are thus lead to the effective Lagrangian
\begin{eqnarray}
{\cal L}(x)=-
\phi^\dagger(x)\left[-(\partial_\mu + i
B_\mu(x))^2+m^2\right]\phi(x), \ B_\mu=\frac{1}{2}B_{\mu\nu}x_\nu,
\nonumber
\end{eqnarray}
and work, initially at least, in Euclidean space.
Because we seek to approach the Wilson criterion,
we consider the analogous Green's function describing
a particle-antiparticle loop. Thus
the object we are interested in is the four-point function
\begin{equation}
\label{loop1}
G(x,y|B)=\langle
:\phi^\dagger(x)\phi(x)::\phi^\dagger(y)\phi(y):\rangle_B=
S(x,y|B)S(y,x|B).
\end{equation}
The normal ordering is taken to exclude the disconnected diagram.
The two-point function $S(x,y|B)$ is itself a solution to the
equation
\begin{eqnarray}
\left[-(\partial_\mu +
iB_\mu(x))^2+m^2\right]S(x,y|B)=\delta(x-y).
\nonumber
\end{eqnarray}
The propagator in the external field transforms under
translations
($x\to x+a$, $y\to y+a$) as
\begin{eqnarray}
\label{tr}
S(x,y|B)=e^{\frac{i}{2}x_\mu B_{\mu\nu}a_\nu}S(x+a,y+a|B)
e^{-\frac{i}{2}y_\rho B_{\rho\sigma}a_\sigma}.
\end{eqnarray}
The Green's function Eq.(\ref{loop1}) is gauge invariant
and, hence, translation invariant.
By means of transformation Eq.(\ref{tr}) with $a=-(x+y)/2$
we rewrite the function Eq.(\ref{loop1}) in a manifestly
translation invariant form:
\begin{eqnarray}
G(x,y|B)=G(x+a,y+a|B)=
G\left((x-y)/2,(y-x)/2|B\right)=W(x-y|B).
\nonumber
\end{eqnarray}
Using the proper time method the propagator
can be represented in the form
of a path integral over a one-dimensional field $\xi$~\cite{Fey},
\begin{eqnarray}
\label{pot2}
 S(x,y|B)=e^{\frac{i}{2}x_\mu B_{\mu\nu}y_\nu}
\int\limits_0^\infty d\alpha {e^{-{\alpha\over2}m^2}
\over8\pi^2\alpha^2}
\int D\xi\exp\left\{-\int\limits_0^\alpha d\tau
\frac{1}{2}\left[\dot{\xi}^2(\tau)+
i\dot{\xi}_\mu(\tau)B_{\mu\nu}\xi_\nu(\tau)
\right]\right\}
\end{eqnarray}
with the boundary conditions $\xi(0)=-(x-y)/2$,
$\xi(\alpha)=(x-y)/2$,
and the
normalisation
$$\int D\xi\exp\left\{-\int\limits_0^\alpha
d\tau\frac{\dot{\xi}^2(\tau)}{2}
\right\}=\exp\{-(x-y)^2/2\alpha\}.$$
Let us first review the confining properties of these
fields in terms of analytical properties of the propagator.
It is instructive to consider first the case of arbitrary constant
$B_{\mu\nu}$.
Since vectors $\vec H\pm \vec E$
($H_i=\epsilon_{ikj}B_{kj}/2$, $E_i=B_{i4}$) are rotated independently
of each other under Euclidean O(4) transformations, the tensor
$B_{\mu\nu}$ can be put into the configuration $B_{34}=E$,
$B_{12}=H$, $B_{13}=B_{14}=B_{23}=B_{24}=0$, and
$H>0$, $-H\le E\le H$~\cite{Leu81}.
The path integral in Eq.(\ref{pot2}) can be easily performed
with the result
\begin{eqnarray}
&&S(x,y|B)=e^{\frac{i}{2}x_\mu B_{\mu\nu}y_\nu}
\frac{H|E|}{16\pi^2}\int\limits_0^\infty
\frac{d\alpha e^{-m^2\alpha}}{\sinh(H\alpha)\sinh(|E|\alpha)}
\nonumber\\
&&\times\exp\left\{-\frac{1}{4}H[(x_1-y_1)^2+(x_2-y_2)^2]\coth(H\alpha)
\right.
\nonumber\\
&&\left.
\ \ \ \ \ \ \ \ \ \
-\frac{1}{4}|E|[(x_3-y_3)^2+(x_4-y_4)^2]\coth(|E|\alpha)\right\}.
\nonumber
\end{eqnarray}
This leads to a Fourier transform of the
translation invariant part
\begin{eqnarray}
\label{eh}
&&\tilde S(p|B)=
\int\limits_0^\infty
\frac{d\alpha e^{-m^2\alpha}}{\cosh(H\alpha)\cosh(|E|\alpha)}
\nonumber\\
&&\times\exp\left\{-\frac{1}{H}(p_1^2+p_2^2)\tanh(H\alpha)
-\frac{1}{|E|}(p_3^2+p_4^2)\tanh(|E|\alpha)\right\}.
\end{eqnarray}
When $E$ is nonzero this function is finite for any complex $p_1^2+p_2^2$
and $p_3^2+p_4^2$ and thus is an entire analytical function.
When $E=0$ this representation exhibits a pole in the physical region
$p_4^2=-(p_3^2+m^2+H)$, which corresponds to a free propagation along
the third axis with the energy equal to the lowest Landau level
of spinless particle. In the $(1-2)$ plane the particle is confined.

Thus for $E \neq 0$, no physical particle corresponding to
the field $\phi(x)$ can appear in the spectrum.
The charged particles are, in other words, confined.
However, as has been shown in \cite{Leu81}, such an abelian constant
field is unstable against small quantum fluctuations
{\it except in the case that it is self-dual or antiself-dual}:
$H=B, E=\pm B$.  In the following, we concentrate precisely
on this configuration. In this case Eq.(\ref{eh}) takes the simple form
($t=\tanh(B\alpha)$)
\begin{equation}
\label{pot22}
\tilde S(p^2|B)=\frac{1}{B}\int\limits_0^1 dt
\left(\frac{1-t}{1+t}\right)^{m^2/2B}\exp\left\{-\frac{p^2}{B}
t\right\}
\end{equation}
which represents an entire function in the complex $p^2$ plane.
A special case is that of $m=0$:
the Fourier transform of the massless propagator turns out to be
\begin{eqnarray}
\label{zerom}
\tilde S(p^2|B)|_{m=0} = \left(1-e^{-p^2/B}\right)/p^2 \ .
\end{eqnarray}
This is manifestly an entire analytical function in the complex
$p^2$-plane: the apparent massless pole at $p^2=0$ simply
cancels out illustrating most cleanly the confinement property.
As a matter of fact, entire propagators mean
that the quantum field theory is nonlocal. It should be noted
here that at the axiomatic level nonlocal quantum field theory
was successfully constructed some time ago \cite{Efi75,FS77,Mof90}.
In particular, causality and unitarity of the $S-$matrix was
proved, a procedure for canonical quantization of nonlocal
field theories was constructed and, recently, Froissart type
bounds on cross-sections at high energy were obtained~
\cite{Efi97}. But to summarise this brief review of known results
for constant fields, we can say that confinement in the sense of
entire propagators is a property of any
Euclidean abelian constant field configuration
with non-zero magnetic {\it and} electric components, but the
(anti-)self-dual case is distinguished by being stable against
quantum fluctuations.

To see how this property can relate to the Wilson criterion,
we now approach the problem of static charges.
We consider heavy particles, with $m^2 \gg B$.
In this limit Eq.(\ref{pot2}) can be represented in the form
of a quantum mechanical path integral (see Appendix A)
\begin{equation}
\label{lag1}
S(x,y|B) \propto e^{-mT}\int D\vec\eta
\exp\left\{-\int\limits_0^Td\beta L(\eta(\beta))\right\},
\end{equation}
where
\begin{eqnarray}
&& L=\frac{m\dot{\vec\eta}^2}{2}-\frac{i}{2}\vec
B[\dot{\vec\eta}\times\vec\eta]+
\frac{1}{2m}({\vec\eta} \cdot {\vec E})^2,
\nonumber\\
&& T=x_4-y_4, \ \ \vec\eta(0)=-(\vec x-\vec y)/2, \ \
\vec \eta(T)=(\vec x-\vec y)/2.
\nonumber
\end{eqnarray}
Here, $E_j=B_{4j}$ is the electric component of the tensor
$B_{\mu\nu}$, and $B_i=-{1\over2}\epsilon_{ijk}B_{jk}$ is the
magnetic component. We will implement the (anti-)self-duality
condition $E_j=\pm B_j$ below. For the present, we insert the
representation Eq.(\ref{lag1}) into Eq.(\ref{loop1}), introduce
the center of mass coordinates
${\vec R}=({\vec\eta}_1+{\vec\eta}_2)/2$,
${\vec r}={\vec \eta}_1-{\vec \eta}_2$,
${\vec R}(0)={\vec R}(T)=0$,
${\vec r}(0)=-{\vec r}(T)={\vec y} - {\vec x}$,
and integrate out the center of mass coordinate $\vec R$. The
integral over $\vec R$ obviously
does not depend on $\vec x$ and $\vec y$, which is simply a
consequence of the translation invariance of the function $W$.
After continuation to physical time $(T= iT,~~\beta=it)$
the result for $W$ is
\begin{eqnarray}
W(\vec x-\vec y,T|B) & \propto & e^{-2i m T}\int D\vec r
\exp\left\{i\int\limits_0^TdtL(\vec r(t))\right\},
\nonumber\\
L&=&{\mu\dot{\vec r}^2\over2}+
{1\over4}\dot{\vec r}[\vec r\times\vec B]-
{1\over8\mu}(\vec{r}\vec{B})^2,
\nonumber
\end{eqnarray}
where $\mu=m/2$ is the reduced mass of the two-particle system.
One sees that the conjugate momentum and the Hamiltonian are
\begin{eqnarray}
\vec{p}&=&\mu\dot{\vec r}+{1\over4}[\vec r\times\vec B]
\nonumber \\
H&=&{{\vec p}^2\over2\mu}
-{1\over4\mu}\vec{p}[\vec{r}\times\vec{B}]
+{1\over32\mu}[\vec{r}^2\vec{B}^2+3(\vec{r}\vec{B})^2],
\label{ham}
\end{eqnarray}
and that the function $W$ can be reexpressed as a phase-space
functional
integral,
\begin{equation}
\label{ham1}
W(\vec x-\vec y,T|B)\propto e^{-2imT}\int D\vec r D\vec p
\ \exp\left\{-i\int\limits_0^Tdt\left[H(\vec r,\vec p)-
\vec p \cdot \dot{\vec r}
\right]\right\}.
\end{equation}
Eqs. (\ref{ham}) and (\ref{ham1}) show that the
massive charged particle and anti-particle in the external
self-dual field
are bounded by an oscillator potential.
Now, consistent with Wilson \cite{Wil74},
we extract from the path integral the contribution to the phase
space
of the stationary trajectory $(\vec p=0, |\vec r|=|\vec x -\vec y|
)$.
Equation (\ref{ham}) indicates that this trajectory corresponds
to uniform circular movement of the particle-antiparticle pair
on a circle with radius $|\vec x-\vec y|$ in the plane perpendicular
to the direction of field $\vec B$.
We find that the contribution is exponentially
suppressed,
\begin{equation}
 \exp\left(-iT\frac{B^2}{32\mu}(\vec x -\vec y)^2\right).
\label{suppression}
\end{equation}
The Wilson criterion for confinement is indeed satisfied.
However here we have a `volume law' rather than an area law.
The relationship between this result and that in standard lattice
gauge theory will be discussed in the final section. For now, 
we stress that the confining potential has appeared
due to the background field, and not due to
interaction between particles via gauge boson exchange.
Such effects will generate additional potential terms to the
Hamiltonian, and will thus affect the energy spectrum of the system. But
gauge boson exchange will not change the basic confining properties of
the background field.

This picture of bound state formation seems strange at first
sight.
However,
an analogy with the quantum dots (or artificial atoms) of solid
state physics can be recognised~\cite{Qdot}.
Quantum dots are quasi-zero-dimensional electron systems
in semiconductor nanostructures in which
three-dimensional confinement of small numbers of electrons
is achieved by a combination of band offsets and electrostatic
means.
The simplest model
Hamiltonian for the few-electron quantum dot
was obtained by solving the Schr\"odinger and Poisson
equations self-consistently within the Hartree
approximation~\cite{Kum90}.
It was found that the oscillator confining potential has nearly
circular symmetry.
The difference in our case is the origin of the confining
potential.
The Hamiltonian Eq.(\ref{ham}) has appeared due
to the background gauge field which may
arise in the vacuum as a result of gluon self-interactions.

In QCD, this picture of confinement and bound state
formation in the static quark limit will be basically the same.
Thus Eqs.(\ref{ham},\ref{ham1}) give illustrative insight into
the basic nature of confinement provided for by the self-dual
field.
But, as Eq.(\ref{zerom}) indicates, the significance of the
property of entireness of Green's functions as a characterisation
of
confinement applies to
dynamical fields and thus is relevant to the fully relativistic
bound state spectrum of QCD, the physically relevant problem.
Thus the qualitative basis for investigation into
the impact of confinement on the relativistic bound state
spectrum are equations like Eqs.~(\ref{pot22}) and (\ref{zerom}),
as has been carried
out in \cite{selfdualphen}. Here an effective meson theory based
on the
bosonisation of nonlocal quark currents has been developed.
The background field has been taken into account both
in quark and gluon propagators.
Within this effective theory the ground and excited state spectra
of
light, heavy-light  mesons and heavy quarkonia have been
calculated, with the only parameters being quark masses, the
background field strength and the gauge coupling constant.
Agreement with experimental data is obtained to within ten
percent.
Regge behaviour within this approach is recovered precisely by
the fact
that
gluon and quark propagators are entire functions. The
relationship
between this mechanism of confinement and flavour chiral symmetry
breaking
is analysed in \cite{EfN98}.

Having explored again the confining properties of the self-dual
homogeneous background field in QCD, we now turn to
the problem of the effective potential for this field
at finite temperature and the question of deconfinement.

\section{Self-Dual Field and Finite Temperature}
In this section we compute the one-loop effective
potential for the self-dual background field at finite
temperature in SU(2) Yang-Mills theory.
This enables us to study its presence or absence
at high temperatures where perturbation theory should become
reliable.

Since we are already in Euclidean space in order to
define the self-dual field, it is convenient to introduce
finite temperature $T$ by working in the imaginary time
formalism. The $x_4$ direction is now a finite interval
of length $\beta = 1/T$ and boundary conditions must be
imposed on the gluon fields to which we shall come below.
We work in a completely gauge-fixed formalism within
which we will introduce the external field.
At zero temperature, the background gauge is most convenient.
However in the present case,
the breaking of manifest Lorentz invariance (by the heat bath)
suggests the temporal (axial) gauge is a natural gauge choice.
Specifically, we choose
\begin{equation}
\partial_4 A_4^a(x) = 0
\label{tempgauge}
\end{equation}
followed by a diagonalisation of the surviving zero mode
$a_4^a({\vec x}) \tau^a = \frac{1}{\beta} \int_0^\beta A^a_4(x)
\tau^a
dx_4$.
This gauge is a special case of the static temporal gauge.
Here one encounters the problem of the nontrivial Haar measure
in the functional integral quantisation of the theory
\cite{Weylgauge}.
Concomitantly, the diagonalised variable $a_4^{\rm diag}({\vec
x})$ is
compact. The functional integral over this variable is thus
non-Gaussian.
Progress on the computation of this integral for SU(2) Yang-Mills
theory
was made recently in \cite{LT98}
wherein, using a lattice regularisation, it was shown
that the integration out of $a_4^{\rm diag}({\vec x})$
leads to an effective action for the remaining degrees of
freedom.
In the absence of external fields, the key features of
this effective theory were that off-diagonal, namely
charged, components of the gluon fields acquired a
temperature dependent mass $M(T)$. Secondly, the boundary
conditions
in $x_4$ of these fields were changed from periodic to
antiperiodic.

We rederive this effective theory in Appendix B, and show that
the presence of the self-dual background field does
not force major modifications. In particular, the
rigorous result for the mass, expected to be valid at low
but non-zero temperatures \cite{LT98,EKLT98}
is reproduced even in the presence of the homogeneous field,
namely,
\begin{equation}
M(T) = \sqrt{(\pi^2/3 - 2)} T
\;.
\label{tempmass}
\end{equation}
In \cite{EKLT98} it was argued that stability
with respect to chromomagnetic fluctuations mean that the
mass term in the deconfined phase should take the form
\begin{equation}
M(T) = {\frac{11}{12\pi}} T g^2(T) , \ T \rightarrow \infty \ ,
\end{equation}
where $g(T)$ is the perturbative running coupling constant.
The important consequence of this result is that at high
temperature the
mass
itself diverges but the ratio $M(T)/T$ vanishes in this limit.
This latter property is sufficient to guarantee the recovery of
the Stefan-Boltzmann law in the high-temperature regime.

Now we consider the self-dual external field and choose it
to point in the same colour direction as $a_4^a \tau^a$.
It is important to note that this corresponds to a distinct
physical choice
since gauge freedom does not allow both $B_\mu^a \tau^a$ and
$a_4^a \tau^a$ to be simultaneously diagonal.

We come to the question of the gluonic boundary conditions.
Here care is required as, unlike the chromomagnetic
choice \cite{Sav77,ElS86,EKLT98}, the
self-dual field involves a component pointing in the,
now compact, time direction. We are therefore no longer
free to impose the usual periodic boundary condition.
Instead, the choice must be consistent now with parallel
transport in the presence of an external field.
Specifically, the appropriate boundary condition in the spatial
directions ${\vec x}$ is the usual vanishing one. For the
direction $x_4$ which is finite, $x_4 \in [0,\beta]$,
one usually chooses periodic boundary conditions in the
absence of external fields. This can be represented in
the form
\begin{equation}
e^{\beta \partial_4} A^a_\mu(x_4,{\vec x}) = A^a_\mu(x_4,{\vec
x})
\; .
\end{equation}
In the presence of an external field $B_\mu^a$ the natural
generalisation
of this for the fluctuating gauge fields
$Q^a_\mu$ is obtained via parallel transport, namely
\begin{eqnarray}
\left( e^{\beta D_4} \right)^{ab}
Q^b_\mu(x_4,{\vec x}) & = & Q^a_\mu(x_4,{\vec x}) \nonumber \\
D^{ab}_4 & = & \delta^{ab} \partial_4 - \epsilon^{3ab} B_4
\label{qperiod}
\; .
\end{eqnarray}
This boundary condition will, in the simplest way, preserve
the periodicity of observable, gauge invariant quantities.
We shall refer to this position-dependent twisted boundary
condition
as {\it quasiperiodic}. When the considerations of Appendix B
are carried out and the zero mode of the {\it fluctuating} gauge
field,
$q_4^{\rm diag}$, is integrated out, Eq.(\ref{qperiod}) becomes
a
quasi-{\it anti}periodic boundary
condition:
\begin{equation}
\left( e^{\beta D_4} \right)^{ab}
Q^b_\mu(x_4,{\vec x})  =  - \ Q^a_\mu(x_4,{\vec x})
\label{qaperiod}
\; .
\end{equation}
To summarise what will be important then for the following
calculation,
there are two key features: firstly, that boundary conditions
are modified to being quasi-antiperiodic, and secondly that the
off-diagonal gluon components have a temperature dependent mass
$M(T)$ which diverges as $T$ increases.
It is precisely this which
gives us a well-controlled high temperature regime
specified by $T \gg \Lambda_{\rm SU(2)}$ and
$B < T^2$.

To calculate the effective potential now,
it is convenient to bring the field-strength tensor to the form
(taking the field $\vec B$ to be directed along the third spatial
axis)
\begin{eqnarray}
\label{str}
  \left(B_{\mu\nu}\right)_{\mu,\nu=1,2,3,4}=
        \left(\matrix{ 0      & -B       & 0     & 0     \cr
                       B      &  0       & 0     & 0     \cr
                       0      &  0       & 0     & \pm B \cr
                       0      &  0       & \mp B & 0
}\right),
\end{eqnarray}
where the upper (lower) sign corresponds to the self-dual
(anti-self-dual)
field. The effective potential is defined in the usual way
using the functional integral,
\begin{equation}
\label{repr1}
 {\cal Z}  =
{\cal N}\int DQ
\exp \left\{ \int d^4x {\cal L}_{\rm eff}[Q_i^A,Q_i^3,B_\mu^3]
\right\}=
\exp\left\{- \beta V U_{\rm eff}(B,\beta,g)\right\}
\end{equation}
where $i,j,k,l=1,2,3, \ A,B = 1,2$ denote spatial and
off-diagonal
field components for gluons respectively, $V$ is the
three-dimensional
spatial volume, and, as derived in Appendix B, the
effective Lagrangian can be written as
\begin{equation}
{\cal L}_{\rm eff} [Q^A_i,Q^3_i,B_\mu^3]
= {\cal L}_{\rm YM} [Q_\mu, B_\mu]|_{Q_4 = 0}
- \frac{1}{2} M^2(T) Q^A_i Q^A_i,
\end{equation}
with ${\cal L}_{\rm YM}$ the standard Yang-Mills action.
The functional integral is defined on the space of
quasi-antiperiodic fields satisfying Eq.(\ref{qaperiod}).
The normalisation in Eq.(\ref{repr1}) is chosen so that
$U_{\rm eff}(0,\beta,g)=0$. To the action, a gauge-fixing
term involving the neutral zero mode gluons
$Q^3_i({\vec x}) = \frac{1}{\beta} \int_0^\beta dx_4 Q^3_i(x)$
can be added, but which decouples at one-loop
after the normalisation at zero field, $B=0$.
Dropping terms in the Lagrangian higher than quadratic in
the fluctuating fields $Q^A_i$, we can extract from ${\cal
L}_{\rm YM}$
the following piece relevant for the one-loop effective potential
\begin{equation}
\label{terms}
{\cal L}  = -\frac{1}{2}
Q_i^A(x)\left[
-(\nabla^2)^{AB}\delta_{ij} + M^2(T)\delta^{AB}
\delta_{ij}+\left(D_i D_j\right)^{AB}
+2B_{ij}\varepsilon^{3AB}
\right]Q_j^B(x)
\end{equation}
where $\nabla^2=D_4^2+D^2_i$. The quadratic operator in
Eq.(\ref{terms})
has zero modes for the case $M=0$ which are called
chromons~\cite{Leu81}.
A correct calculation of the chromon
contribution to the effective potential at zero temperature
requires
an extension of the one-loop approximation:
an interaction (or mixing) between zero modes and normal modes
has to be taken into account.  However, we take the
temperature to be sufficiently large so that $M^2(T)$ is
correspondingly
large
compared to $B$. Large $M$ means that the contribution of
chromons is
regular
at one-loop order so the mixing between them and normal modes can
be
neglected. The one-loop effective potential is thus given by
\begin{equation}
\label{p-eff}
U_{\rm eff}=\frac{B^2}{g_0^2}+\frac{1}{V \beta}{\rm Tr}\ln\left[
\frac{\left(-\nabla^2\delta_{ij}+D_i D_j
-2iB_{ij}+M^2(T) \delta_{ij}\right)}
{\left(-\partial^2\delta_{kl}+\partial_k \partial_l
+M^2(T) \delta_{kl}\right)}\right].
\end{equation}
Here $B^2/g_0^2$ comes from the classical action with $g_0$ is
the bare
coupling constant. The effective potential can be rewritten in
the form
\begin{equation}
\label{p-eff1}
U_{\rm eff}=\frac{B^2}{g_0^2}-
\int_V \frac{d^3x}{V}
\int\limits_0^{\beta} \frac{dx_4}{\beta}
\int\limits_{M^2(T)}^\infty d m^2\left[
D_{ii}^{\beta}(x,x|B,m)-D_{ii}^{\beta}(x,x|0,m)
\right].
\end{equation}
We have to calculate the trace of the propagator
$D_{ij}^{\beta}(x,y|B,m)$ satisfying
quasi-antiperiodic boundary conditions, Eq.(\ref{qaperiod}):
\begin{eqnarray}
\label{eq3}
D_{ij}^\beta(x_4+\beta ,{\vec x};y|B,m) & = &
-D_{ij}^\beta(x_4,{\vec x};y|B,m)
\exp[-i\beta B_4({\vec x})] ,\nonumber \\
D_{ij}^\beta(x;y_4+\beta,{\vec y}|B,m) & = &
-D_{ij}^\beta(x;y_4,{\vec y}|B,m)
\exp[i\beta B_4({\vec y}) ].
\end{eqnarray}
This can be implemented by first solving for the
Green's function $\Delta_{ij}(x|B,m)$ relevant to the zero
temperature or infinite volume
and then building up the Green's function satisfying the
finite temperature boundary condition via
(see also \cite{Mak94} and references therein)
\begin{eqnarray}
\label{l-prop}
D_{ij}^{\beta}(x;y|B,m) & = &
\sum\limits_{n=-\infty}^{\infty}(-1)^n
\Delta_{ij}(x_4-y_4+n \beta;{\vec x}-{\vec y} \ |B,m)
\nonumber\\
&{}&
 \times\exp\left(\frac{i}{2}x_\mu B_{\mu\nu}y_\nu +
\frac{i}{2}n \beta B_4({\vec x} + {\vec y})\right).
\end{eqnarray}
It should be stressed that Eq.~(\ref{l-prop}) implies the
existence
of an orthogonal complete set of eigenfunctions of the operator
$\nabla^2$ satisfying the quasi-antiperiodic boundary conditions.
The existence of such a set of functions is demonstrated in
Appendix~C.

The infinite volume or zero temperature
Green's function $\Delta_{ij}$ is a solution to the equation
\begin{eqnarray}
%\label{green2}
\left[\left(-\nabla^2+m^2\right)\delta_{ij}+D_i D_j
-2iB_{ij}\right]
\Delta_{jk}(x|B,m)=\delta_{ik}\delta(x).
\end{eqnarray}

A complete solution of this system of equations is quite
involved,
but the trace of the propagator is tractable as is shown in
Appendix~D.
One comment is in order though: the summation over $n$ in the
space-time trace of Eq.(\ref{l-prop})
is suppressed in the infinite volume limit $V \rightarrow \infty$
due to the electric field component of the self-dual field.
So in fact the only relevant contribution of Eq.(\ref{l-prop})
to the
effective
potential Eq.(\ref{p-eff1})
is that from $n=0$. Using this fact and results
Eqs.~(\ref{system3}) and (\ref{general}) derived in Appendix D,
we arrive at the relation
\begin{eqnarray}
\label{tr1}
\Delta_{ii}(0|B,m) & = & \sum_k\left[F_k(x,x)
-\int d^4z F_k(z,x)\tilde D^2_k(x)\Delta_4(x,z) \right]
\nonumber \\
& {} & + 2B^2\int d^4z\int d^4z^\prime
F_0(x,z)\Delta_4(z,z^\prime)
F_0(z^\prime,x),
\end{eqnarray}
where
\begin{eqnarray}
\label{fk}
F_k(x,y) & = &\exp\left(\frac{i}{2}x_\mu B_{\mu\nu}y_\nu\right)
\frac{B^2}{16\pi^2}\int_0^\infty
\frac{dr}{\sinh^2(Br)}
\nonumber \\
&{}& \times
\exp\left[-m^2r+2B\xi_kr-\frac{1}{4}(x-y)^2B\coth(Br)\right],
\\
\Delta_4(x,z) & = & \frac{1}{2\sqrt{\pi}}\delta^{(3)}
({ \vec x} -{\vec z})
\int_0^\infty\frac{dt}{\sqrt{t}}
\exp\left[-m^2t-\frac{(z_4-x_4)^2}{4t}
-\frac{i}{2}(z_4-x_4)B_{4j}x^j \right].
\nonumber
\end{eqnarray}

According to Eqs.~(\ref{p-eff1}), (\ref{tr1}) and (\ref{fk}),
the effective potential can be expressed as the combination
\begin{equation}
\label{ueff-f}
U_{\rm eff}(B^2)  =
\frac{B^2}{g_R^2(M)}+U_1(B^2)+U_2(B^2)+U_3(B^2),
\end{equation}
where
\begin{eqnarray}
U_1 & = & -\frac{B^2}{16\pi^2}\int_0^\infty\frac{ds}{s^3}
\exp\left(-\frac{M^2}{B}s\right)
\left\{\frac{s^2}{\sinh^2s}\left[1+2\cosh(2s)\right]-3s^2-3
\right\},
\nonumber \\
U_2 & = & -\frac{B^2}{32\pi^2}\int\!\!\int_0^\infty\frac{ds
dt}{s^2(s+t)}
\exp\left(-\frac{M^2}{B}(s+t)\right)
\nonumber\\
&{}& \times \left\{\frac{s^2}{\sinh^2s}\cdot
\frac{2\sinh(2s)-\coth s\left[1+2\cosh(2s)\right]}
{\sqrt{1+t\coth s}}
\right.
\nonumber\\
&{}&
\left.
+\frac{3}{\sqrt{s(s+t)}}-\frac{ts^2}{2(s+t)\sqrt{s(s+t)}}
\right\},
\nonumber\\
U_3 & = & -\frac{B^2}{8\pi^2}\int\!\!\int\!\!\int_0^\infty
\frac{ds dr dt}{(s+r+t)}
\exp\left(-\frac{M^2}{B}(s+r+t)\right)
\nonumber \\
&{}& \times
\left\{
\left[\sinh( s+ r)\right]^{-3/2}\left[\sinh( s+ r)
+t\cosh( s- r)\right]^{-1/2}
\right.
\nonumber\\
&{}&
\left.
-(s+r)^{-3/2}(s+r+t)^{-1/2}\right\}.
\end{eqnarray}
The functions $U_1$, $U_2$ and $U_3$ correspond to
ultraviolet finite contributions of the three terms
in Eq.~(\ref{tr1}). The renormalised coupling constant
$g_R^2$ runs with the scale defined by the mass $M=M(T)$
and is
\begin{equation}
\frac{1}{g_R^2}=\frac{1}{g_0^2}\left(1-
b_0\int_{s_0}^\infty\frac{ds}{s}
e^{-M^2s}\right),
\nonumber
\end{equation}
where we have used (gauge invariant) Schwinger regularisation.
The renormalization procedure we use corresponds to the
zero momentum subtraction scheme.
Taking the parameter $s_0\to 0$ generates the ultraviolet
divergence.
The constant $b_0$ is nothing but the coefficient of the beta
function
to lowest order. Its value arises as a sum of the divergent parts
of the
three terms  in Eq.(\ref{tr1}) which give for
$U^{\rm div}_1$, $U^{\rm div}_2$ and $U^{\rm div}_3$
contributions 3/16, 1/48 and 1/4, respectively, so that
$$
b_0 = 11/24
$$
correctly arises. That we get this is another check on
the consistency of our formalism. In particular, by renormalising
in this
way
we have combined the ${\cal O}(B^2)$ quantum corrections
with the classical term, so that the next corrections begin
at ${\cal O}(B^4)$.

We thus obtain our final result for the effective potential
at high temperature:
\begin{equation}
 U_{\rm eff}(B^2) = {{B^2} \over {g^2_R(M)}} +
{{29}\over{525\pi^2}} {{B^4} \over {M^4(T)}}
+ {\cal O}(B^6/M^8(T)) + {\cal O}(g^2_R(M))
\ .
\label{potresult}
\end{equation}
Since, as $T \rightarrow \infty$, $M(T) \rightarrow \infty$
we have $g_R(M) \ll 1$, and our calculation is reliable in this
regime. So the effective potential
acquires a minimum at zero value
of the external field. The background field switches
off at high temperature, and we can characterise the high
temperature phase
as exhibiting deconfinement.

\section{Discussion}
The central results of this work are expressed
in Eqs.(\ref{suppression},\ref{potresult}). From these we
understand that if confinement is due to fields in the QCD vacuum
which are long-range (homogeneous in our case) and satisfy
self-duality or anti-self-duality then this is
neither in conflict with the Wilson criterion for static quark
systems nor with the natural expectation that with increasing
temperature there is a transition from confinement to
deconfinement. It will be immediately noticed that
we have not obtained a linear heavy quark potential as has
been observed in lattice simulations. The reason for this
discrepancy is straightforward: lattice calculations normally
implement periodic boundary conditions from the very outset.
As shall be repeatedly seen in the following, the existence of
fields non-vanishing at infinity entails significant problems for
incorporation on the lattice due to the quasiperiodic boundary
conditions. It seems that there is strong evidence that
lattice calculations of the heavy quark potential have {\it quite
correctly}
not seen a quadratic potential because the effects of the vacuum
field we consider have not been built in. How to build these
effects in is a problem we discuss below.

On the other hand, the self-dual field, at least at the
level of the lowest order propagators in this background, does not
immediately account for all aspects of confinement:
diagonal gluons in SU(2) have poles
in the propagator and this is a consequence of the fact that
they do not couple directly to the diagonal background
configuration.  As mentioned in the introduction, the simple
self-dual homogeneous configuration
is not the entire story, and there is room for local effects
which can complete the picture of confinement.
The vacuum field breaks spontaneously CP, colour and
O(3) symmetries. There is a continuum of degenerate vacua
corresponding to different directions of the vacuum field.
This implies~\cite{Sw} the existence of soliton-like field configurations
under the homogeneous background field,
which could play the role of topologically nontrivial local
defects in the QCD vacuum such as domain walls.
In the absence of explicit solutions we can only speculate
on the robustness of our results against inclusion of such effects.
But insofar as the confining properties of the self-dual
homogeneous field depend only on the strength of the field
and not the direction (in real and internal space), it seems plausible
that domains distinguished only by changes in direction will not
disrupt the confinement we observe.

However, all of this rests on the assumption that
at zero and low temperatures
the effective potential for this background has a minimum at
non-zero field value and there is no substitute for a genuine
nonperturbative calculation. The only realistic choice for
this is the formalism of lattice QCD. We thus discuss now in
some detail the problems to be confronted with setting up the
calculation on the lattice, and some insights our preliminary
investigation into this offers.

The essential question we need to answer is what
the contribution of the homogeneous field configuration
Eq.(\ref{field}) is to the partition function of lattice $SU(2)$
gauge theory
\begin{equation}
\label{lat1}
Z=\int\limits_{\cal U}DU\exp\left\{-S[U]\right\}.
\end{equation}
Here, $S$ is now the standard Wilson action and $U$ is shorthand
for
\begin{equation}
\label{lat2}
U_{n,\mu}=\exp\left\{iaA_{\mu}(an)\right\}\in SU(2), \ \ \forall
n,\mu,
\end{equation}
the link variable, and $DU$ is a functional Haar measure.
The lattice spacing is $a$. Link variables
are functions of $n$ and are subject to some boundary condition.
They thus belong to some functional space $\cal U$.
Usually, with $N$ representing the size of the lattice in
a given direction, periodic boundary conditions
$$
U_{n+N,\mu}=U_{n,\mu}
$$
are imposed in order to implement the translation invariance of
the theory
in the thermodynamic limit. However, the field
$B_\mu(an)=aB_{\mu\nu}n_\nu$
is
evidently not translation-invariant.
In principle, there are two ways to proceed, both of which have
been used in application
to the Savvidy chromomagnetic background \cite{Sav77}.
The first choice is to force the
long range modes to be simply periodic on the lattice. This can
be done by
`quantisation' of the field strength \cite{CeC97,CeC91,AMBZ89}
\begin{equation}
a^2B_{\mu\nu}=2\pi\frac{b_{\mu\nu}}{N},
\label{discfield}
\end{equation}
where the matrix elements $b_{\mu\nu}$ are integers.
This certainly provides for  periodicity of the corresponding
link
variable,
but rewriting Eq.(\ref{discfield}) as
$$
B_{\mu\nu}=2\pi\frac{b_{\mu\nu}}{aL}, \ \ L=aN,
$$
and going to the thermodynamic ($L\to\infty$) continuum ($a\to
0$) limit
one obtains
$$
B_{\mu\nu}=2\pi b_{\mu\nu}/C, \ \ 0\le C \le \infty,
$$
so that the field strength is discretised into multiples of
$2\pi/C$ {\it even in the continuum thermodynamic limit}.
Moreover, the constant $C$ itself depends on our choice of limiting
prescription. These outcomes render this approach rather unappealing.
A second approach is to change boundary conditions.
Free boundary conditions have been advocated in the approach of the authors
of \cite{TrW93} who apply it to the lattice calculation
of the effective potential for the chromomagnetic field in
three-dimensional $SU(2)$ theory.

In our case, Eq.(\ref{qperiod}) suggests the following
generalisation
for link variables. We decompose the general field $A_\mu$ in
Eq.~(\ref{lat2})
into a long range part $B_\mu$ and the fluctuation $Q_\mu$:
$A_\mu(an)=Q_\mu(an)+aB_{\mu\nu}n_\nu\tau^af^a/2$, $f^2=1$.
Quasi-periodic
boundary conditions for the fields $Q$ can be generalised from
Eq. (\ref{qperiod}) to all directions now,
\begin{equation}
\label{lat3}
Q_\mu(a(n+N))=e^{iw(n)}Q_{\mu}(n)e^{-iw(n)}, \ \
w(n)=a^2N_\alpha B_{\alpha\beta}n_\beta\tau^af^a/2.
\end{equation}
Thus the following transformation of link variables is generated
\begin{equation}
\label{lat4}
U_{n+N,\mu}=e^{iw(n)}U_{n,\mu}e^{-iw(n+\mu)}.
\end{equation}
This has the structure of a gauge transformation.  Thus
gauge-invariant quantities such as the action are invariant.
An integral in
Eq.~(\ref{lat1}) includes an integration over all possible values
of the strength tensor $B_{\mu\nu}$ hence all values of $w(n)$,
thus the boundary conditions Eq.(\ref{lat3})
and (\ref{lat4}) are actually free, consistent with \cite{TrW93}.

The most direct step next is to
formulate the effective potential as the lattice functional
integral
$$ \int_{{\cal U}_Q} DU \exp \left\{ - S[U \cdot V] \right\}  $$
with $V$ denoting a link variable generated by the background
field,
$$ V_{n,\mu} = \exp \left\{ i a^2 B_{\mu \nu} n_\nu \tau^bf^b/2
\right\} $$
and where ${\cal U}_Q$ is now the space of
quasiperiodic functions. Here $B_{\mu\nu}$ and $f^a$ are
external and particular directions in the color and euclidean space can
be fixed.
The actual problem is
to find an appropriate representation of the
measure of the integral such that the exclusion of
the given background field is manifest.

The consequence of the gauge function $w(n)$ in
Eqs.~(\ref{lat3},\ref{lat4}) being nonzero over the whole lattice
is that
all degrees of freedom are affected by the gauge transformation.
Thus inclusion of covariantly constant field configurations
in the space of integration $\cal U$ in Eq.~(\ref{lat1})
means actually that the space of allowed gauge
functions cannot be restricted to the class of functions with
local
support.
A significant consequence of this is that Elitzur's theorem
which forbids spontaneous breakdown of local gauge symmetry
\cite{Eli75}
does not apply to this situation. This theorem concerns the
integral
\begin{equation}
\label{lat5}
\lim_{J\to 0}\lim_{N\to\infty}
\langle F(U)\rangle_{N,J}=
\lim_{J\to 0}\lim_{N\to\infty}
Z^{-1}_{N,J}\int\limits_{\cal U}DU\int\limits_{\cal G}DgF(U^g)
\exp\left\{-S[U]+JU^g\right\},
\end{equation}
where $F(U)$ is gauge noninvariant, and $J$ is an external source
which
breaks gauge invariance. The order of limits is important. The
theorem
states
that if gauge transformations $\cal G$ are local -- namely that
they
act on a finite (independent of $N$) number of degrees of freedom
--
then for sufficiently small sources $|\!|J|\!|<\epsilon$ the
following inequality holds
\begin{equation}
\label{ineq}
|\exp\left\{JU^g\right\}-1|\le\eta(\epsilon) \ ,
\end{equation}
with $\eta(\epsilon)$ being independent of $N$ and vanishing as
$\epsilon$
goes to zero. Periodicity of the functions in $\cal U$ is
implicit.

As has been argued above, both conditions exclude
covariantly constant field configurations which are
long range modes that can produce symmetry breaking.
Periodic boundary conditions and locality of gauge functions
are in conflict with a self-consistent incorporation of these
modes
in the lattice functional integral. The choice of
free boundary conditions for $\cal U$
and, in particular, the presence in $\cal G$ of gauge
transformations
which can act on all degrees of freedom results in nonuniformity
in the
function
$\eta$ in lattice size $N$, so that $JU$ becomes an extensive
quantity.
In view of this, the drastic difference in the results of
\cite{TrW93}
(some evidence of nontrivial minimum with free boundary
conditions
and continuous field) and
\cite{CeC97,CeC91} (minimum at zero field strength
with periodic boundary conditions
and `quantised' field strength) seem unsurprising.

It would be instructive to give an example illustrating that inclusion
of the homogeneous fields into the lattice integral allows the existence
of an order parameter that is not gauge invariant.
Let us consider the integral over $\cal U$ and $\cal G$ which includes
now homogeneous fields and gauge functions of the form Eq.(\ref{lat3}).
If we put
$$F(U^g)={\rm Im}U^g_{n,\mu\nu},$$
where $U_{n,\mu\nu}$ is a plaquette variable, and choose the source term
in the form
$$\sum\limits_{n,\mu\nu}J_{\mu\nu}{\rm Tr \ Im}\tau^3U^g_{n,\mu\nu}, \ \
J_{\mu\nu}=const,$$
then the inequality (\ref{ineq}) is not uniform in $N$ for all
field and gauge functions: if $U_{n,\mu\nu}$ contains
the long range fields $a^2 B_{\mu\nu}n_\nu f^b\tau^b/2$
and gauge transformation correspond to
$\omega(n)=a^2N_\alpha B^\prime_{\alpha\beta}n_\beta f^{\prime b}\tau^b/2$,
then one gets for the source term
\begin{eqnarray}
&&\sum\limits_{n,\mu\nu}J_{\mu\nu}{\rm Tr \ Im}
e^{-i\omega(n)}\tau^3e^{i\omega(n)}\exp\{-ia^2B_{\mu\nu}f^b\tau^b\}=
\nonumber\\
&&-2\sum\limits_{n,\mu\nu}J_{\mu\nu}\sin(a^2B_{\mu\nu})
\left[f^3-2(f^{\prime 3}f^bf^{\prime b}-f^3)
\sin^2(a^2N_\alpha B^\prime_{\alpha\beta}n_\beta/2)\right].
\nonumber
\end{eqnarray}
Let for simplicity
$B^\prime_{13}=B^\prime_{14}=B^\prime_{23}=B^\prime_{24}=0$,
$B^\prime_{12}=B^\prime_{34}=B^\prime$, and $N_1=N_2=N_3=N_4=N$.
Using summation formulae
\begin{eqnarray}
&&\sum\limits_{n=1}^{N}\sin^2(nx)=N/2-\cos(N+1)x\sin Nx/2\sin x,
\nonumber\\
&&\sum\limits_{n=1}^{N}\cos^2(nx)=N/2+\cos(N+1)x\sin Nx/2\sin x,
\nonumber\\
&&\sum\limits_{n=1}^{N}\sin(nx)=
\sin\frac{N+1}{2}x\sin\frac{Nx}{2}{\rm cosec}\frac{x}{2},
\nonumber
\end{eqnarray}
one gets in the limit $N\to\infty$
\begin{eqnarray}
&&\sum\limits_{n_1,n_2,n_3,n_4=0}^{N}\sin^2[a^2B^\prime
N(n_1-n_2+n_3-n_4)/2]
\nonumber\\
&&=\frac{N}{2}\sum\limits_{n_2,n_3,n_4=0}^{N}
\{\sin^2[a^2B^\prime N(n_3-n_2-n_4)/2]+\cos^2[a^2B^\prime
N(n_3-n_2-n_4)/2]\}
+O(N^3)
\nonumber\\
&&=N^4/2+O(N^3).
\nonumber
\end{eqnarray}
Thus for the source term we arrive at the result
\begin{eqnarray}
&&-N^4\left( 4f^3-2f^{\prime 3}f^bf^{\prime b}\right)
\sum\limits_{\mu\nu}J_{\mu\nu}\sin(a^2B_{\mu\nu}) + O(N^3),
\nonumber
\end{eqnarray}
which shows that the gauge dependent part of the source term
$JU$ is an extensive quantity, and the order of limits
$J\to0$ and $N\to\infty$ cannot be interchanged.

It should be stressed that this example in no way violates Elitzur's
theorem, but just underlines that its conditions are too
restrictive for a self-consistent incorporation of
homogeneous field configurations into the lattice functional integral
(as mentioned also in the last reference of \cite{Eli75}).

We repeat that the picture of confinement with a
self-dual homogeneous field can become reliable only
with the inclusion of domain structures in the
vacuum such that the symmetries broken by this field
are restored at the hadronic level. The boundaries of the
domains should be describable by some solitonic classical
configurations. As far as we are aware, appropriate solutions are unknown.
It thus remains a problem to verify our considerations
of the Wilson criterion and Elitzur's theorem in
the presence of domains, though we have given plausibility arguments why
our results might be unaffected. The conclusion is thus
that there are two interesting unsolved problems to be
confronted which may be significant for understanding
QCD vacuum structure: a calculation of the effective
potential for the field Eq.(\ref{field}) in the strong coupling
limit, and a search for topologically non-trivial
classical configurations in the background of a
homogeneous self-dual field.

\section*{Acknowledgements}
The authors express their indebtedness
to many intensive and constructive discussions with F.~Lenz,
without whose incisive questions this work would not have
been possible. We would like to thank
L.~von Smekal for interesting discussions.
S.N.N. is grateful to A.~Dorokhov, A.~Efremov, A.~Isaev, 
P.~Minkowski and J.~Moffat for valuable comments.
We acknowledge support
from the Heisenberg-Landau Program for visits
respectively to Dubna and Erlangen. Both institutes
are thanked for kind hospitality
during extensive visits. At the final stage of work S.N.N.
was supported by BMBF grant 06ER809.
G.V.E. was partially supported by RFFR under
grant No.~96-02-17435-a.

\setcounter{equation}{0}
\renewcommand{\theequation}{A.\arabic{equation}}

\section*{Appendix A: Heavy mass limit.}
Below we derive Eq.~(\ref{lag1}) starting with Eq.~(\ref{pot2}).
In the integrand of Eq.~(\ref{pot2}) let us integrate over
$\xi_4$
\begin{eqnarray}
\int D\xi_4\exp\left\{-\int\limits_0^\alpha d\tau
\left[{\dot{\xi}_4^2(\tau)\over2}+
i\dot{\xi}_4(\tau)(\vec{E}\vec{\xi}(\tau))
\right]\right\}=e^{-{T^2\over2\alpha}}
\exp\left\{-\int\limits_0^\alpha d\tau
\frac{1}{2}(\vec{E}\vec{\xi}(\tau))^2\right\}
\nonumber
\end{eqnarray}
where $T=x_4-y_4$ and we have changed variables
$$ \alpha=T/ms, \ \quad \tau=\beta/ms,   \ \quad
\vec{\xi}={\vec{\eta}\over\sqrt{ms}}.$$
Then the propagator takes the form
\begin{eqnarray}
S(x,y|B)\propto \int\limits_0^\infty ds
e^{-{mT\over2}\left(s+{1\over s}\right)}
\int D\vec{\eta}\exp\left\{-\int\limits_0^T d\beta
\left[{\dot{\vec{\eta}}^2\over2}-{i\over 2ms}
\vec{B}[\dot{\vec{\eta}}\times\vec{\eta}]
+{1\over2(ms)^2}(\vec{E}\vec{\eta})^2
\right]\right\},
\nonumber
\end{eqnarray}
where we have omitted the phase factor and a constant
in front of the integral.
One can see that for $T\to\infty$ or more precisely for
$$ {\vert\vec{x}-\vec{y}\vert\over T}\ll 1$$
and $\vert\vec{B}\vert\sim\vert\vec{E}\vert\ll m^2$ the
integral over $s$ can be evaluated using saddle-point
approximation. The saddle-point is $s=1$, hence we
arrive at
\begin{eqnarray}
\label{asp}
 S(\vec{x},\vec{y},T|B)\propto e^{-mT}
\int D\vec{\eta}\exp\left\{-\int\limits_0^T d\beta
\left[\frac{m}{2}\dot{\vec{\eta}}^2-
\frac{i}{2}\vec{B}[\dot{\vec\eta}\times\vec\eta]
+{1\over 2m}(\vec{E}\vec{\eta})^2
\right]\right\}.
\end{eqnarray}
Inserting this result into Eq.~(\ref{asp}), we arrive at the
representation Eq.(\ref{lag1}).

At zero field $\vec E=\vec B=0$
we arrive at the correct nonrelativistic limit
\begin{eqnarray}
&& S(\vec{x},\vec{y},T|B)\propto
e^{-mT-{m\over2}{\vert\vec{x}-\vec{y}\vert^2\over T}}=
\exp\left\{-\left(m+{mv^2\over2}\right)T\right\},\\
&& v={\vert\vec{x}-\vec{y}\vert\over T}\ll1,
\end{eqnarray}
where the energy of
the particle is
$$ E=m+{mv^2\over2}.$$

\setcounter{equation}{0}
\renewcommand{\theequation}{B.\arabic{equation}}

\section*{Appendix B: Static Temporal Gauge and Self-Dual Fields}
Consider a fixed direction in colour
space such that the phase of the gauge invariant Polyakov loop
${\rm Tr P}\exp(ig \int_0^\beta dx_4 A_4^a \tau^a/2)$
is in the $\tau^3$ direction. The background self-dual
field is also chosen to point in the same colour
direction. Gauge transformations now used to fix the gauge
further may not change colour axes.

An arbitrary gauge transformation on the gauge field
${\tilde A}_\mu = {\tilde A}^a_\mu \tau^a/2$
takes the form
\begin{equation}
{\tilde A}_\mu \rightarrow U {\tilde A}_\mu U^{\dagger} +
\frac{i}{g}
U \partial_\mu U^{\dagger} \equiv A_\mu
\ .
\end{equation}
Under the decomposition of ${\tilde A}_\mu$ into
background ${\tilde B}_\mu$ and fluctuating ${\tilde Q}_\mu$
parts
we choose the separate pieces to transform under
`quantum' gauge transformations:
\begin{eqnarray}
B_\mu & = & U {\tilde B}_\mu U^{\dagger} \nonumber \\
Q_\mu & = & U {\tilde Q}_\mu U^{\dagger} +
\frac{i}{g} U \partial_\mu U^{\dagger} \ .
\label{qgauge}
\end{eqnarray}
Since we have specified the direction of the background,
$B_\mu = {\tilde B}_\mu$.

We use the type Eq.(\ref{qgauge}) to achieve
a fixing of the gauge on ${\tilde Q}_\mu$.
These fields satisfy quasiperiodic boundary conditions,
Eq.(\ref{qperiod}),
\begin{eqnarray}
\left( e^{\beta D_4} \right)^{ab} {\tilde Q}^b_\mu(x_4,{\vec x})
& = &
{\tilde Q}^a_\mu(x_4,{\vec x})  \nonumber \\
D^{ab}_4 & = & \delta^{ab} \partial_4 -  \epsilon^{3ab} B_4
\end{eqnarray}
which must not be changed under gauge-fixing.
Thus, $U$ must be quasiperiodic. The temporal
(axial) gauge $Q_4=0$ cannot be achieved with such a group
element.
The static temporal gauge Eq.(\ref{tempgauge})
followed by diagonalisation of the zero mode is however possible.
Explicitly the $U$ bringing an arbitrary ${\tilde Q}_\mu$ into
this gauge is
\begin{equation}
U[{\tilde Q}]
 = e^{- i g x_4 q_4 \tau^3/2} {\rm P} \exp ( i g \int_0^{x_4} dt
{\tilde Q}_4(t,{\vec x}) )
\end{equation}
with $q_4 = {1 \over \beta} \int_0^\beta dt Q_4$.
Since ${\tilde Q}_\mu$ is quasiperiodic so too is $U$.
Thus $Q_\mu$ are also quasiperiodic:
$e^{\beta D_4} Q_\mu(x_4,{\vec x}) = Q_\mu(x_4,{\vec x})$.

There are still two classes of gauge symmetry remaining:
1. Transformations $V(x) = \exp[ i g \omega^3(x) \tau^3/2]$
with $\omega^3(x)$ strictly periodic in $x_4$
can be fixed by introducing an extra Lorentz/Coulomb gauge
condition
on the zero modes of the remaining neutral fields.
As it is Abelian, this gauge fixing does not introduce
Faddeev-Popov
ghosts.
The one-loop effective potential considered in the main body
receives no contributions from these neutral fields.
2. Transformations $W(x) = \exp[(2 i n \pi x_4/\beta) \tau^3 /2]$
cause a shift of $2 n \pi / g \beta$
in the zero mode of the fluctuating field $q_4$.
These are relevant for what follows.

We now implement these considerations in the quantum theory,
using the Faddeev-Popov trick in the functional
integral. The Faddeev-Popov determinant
is defined by
\begin{equation}
\Delta_F^{-1}[Q]  =  \int {\cal D} g \delta[F[Q^g]]
\label{FPdet}
\end{equation}
with $Q^g$ all configurations related
by gauge transformations Eq.({\ref{qgauge}) to a representative
configuration $Q$ which satisfies $F[Q] = 0$.
The functional $F$ that selects this gauge
is independent of the background field $B_\mu$
(unlike in the background field gauge). Inserting unity into
the generating functional we obtain
\begin{equation}
{\cal Z}[B^2] = {\cal N} \int {\cal D} Q_\mu {\cal D} g
\delta[F[Q^g]]
\Delta_F[Q] \exp(- S[B + Q])
\; .
\end{equation}
Now we perform a gauge transformation of type Eq.({\ref{qgauge})
to bring $Q^g \rightarrow Q$. The measure and determinant are
invariant, as stated. To recover the same action,
a corresponding rotation of the background field must take
place, as in Eq.({\ref{qgauge}). Because ${\cal Z}$ is
ultimately a functional only of the gauge invariant
combination $B^2$, we recover again the same ${\cal Z}$.
We may now absorb the integration
$ \int {\cal D} g $ into the normalisation in the usual way and
obtain
\begin{equation}
{\cal Z}[B^2] = {\cal N}  \int {\cal D} Q_\mu \delta[F[Q]]
\Delta_F[Q] \exp(- S[B + Q])
\; .
\end{equation}
The form of the determinant for the static temporal gauget
is well known. Using a lattice regularisation for
space ${\vec x}$, it can be written as
\begin{equation}
\Delta_F[Q] = \prod_{{\vec x}} \sin^2(g \beta q_4({\vec x})/2).
\end{equation}
The Jacobian is independent of the background component $B_4$.
The zeroes of the Jacobian indicate the appropriate range of
integration
for $q_4$, which in turn is
seen in the symmetry under transformations $W$ at the
classical level. The appropriate functional integral
after implementing the delta functional is then
\begin{equation}
{\cal Z}[B^2] = {\cal N}
\int {\cal D} Q_i(x) \int_0^{\pi / g \beta} {\cal D} q_4 ({\vec
x})
 \sin^2(g \beta q_4({\vec x})/2) \exp(-S[B + Q]_{F[Q] = 0})
\; .
\end{equation}
This is still symmetric under the $V$
transformation. Performing the Faddeev-Popov trick
again with the Lorentz/Coulomb gauge condition on the neutral
zero mode
fields enables factoring out of this redundant gauge volume.
With the normalisation ${\cal N}$ being done at $B=0$,
the neutral field contributions to this functional integral
will anyway cancel out. The $W$ symmetry is however fixed
by restriction of the range of integration of $q_4$.

We now show how the integral over $q_4$ can be performed.
We consider
\begin{equation}
\int {\cal D} Q_i {\cal D} q_4 \sin^2(g \beta q_4/2)
\exp\left\{-S+\int d^4x JQ\right\}
\; .
\end{equation}
We integrate over $q_4$ in a diagrammatic expansion in order to
derive an effective theory for $Q_i$.
Thus fields $Q_i$ and $B_\mu$
appear only in external lines of the diagrams.
This is a strong restriction on the allowed diagrams.
The zero mode $q_4$ couples only to charged gluons via the
three- and four-point vertices, and never
to itself. The three-gluon vertex leads to a $q_4 \rightarrow$
two-charged
gluon vertex, while the four-point vertex gives $(q_4)^2
\rightarrow$
charged-anticharged spatial gluons.
This means that the perturbation series in this functional
integral
stops at one loop. Only three topologically distinct classes
of diagrams are present and of these only one specific diagram
gives a non-vanishing contribution as the lattice spacing is
taken zero after subtraction out of a pure infinite constant.
This leaves a mass term in the off-diagonal fields.
Its form is determined by the propagator for
two fluctuating fields $q_4({\vec x})$, namely
\begin{equation}
\langle 0 | T( q_4({\vec x}) q_4({\vec y}) ) |0 \rangle
= { { \int {\cal D} Q_i \int {\cal D} q_4({\vec z})
\sin^2(g \beta q_4({\vec z})/2) e^{ -S_0[B + Q] }
 q_4({\vec x}) q_4({\vec y}) } \over { \int {\cal D} Q_i
\int {\cal D} q_4({\vec z}) \sin^2(g \beta q_4({\vec z})/2)
e^{ -S_0[B + Q] } } }
\label{propdef}
\end{equation}
with $S_0$ representing the action with the couplings between
$q_4$ and the remaining fields dropped.
The unregularised form for this is
\begin{eqnarray}
S_0[B + Q] & = & {L \over 2} \int d^3x (B_4({\vec x}) + q_4({\vec
x}) )
{\vec \nabla}^2 (B_4({\vec x}) + q_4({\vec x}) )
\nonumber \\
& = &
{\beta \over 2} \int d^3x q_4({\vec x}){\vec \nabla}^2 q_4({\vec
x})
\nonumber \\
& = & S_0[Q]
\end{eqnarray}
because
${\vec \nabla}^2 B_4 = \partial_i \partial_i B_{4 j}x_j =
\partial_i B_{4
i}=0$
and $q_4({\vec x}) \rightarrow 0$ as $|{\vec x}| \rightarrow
\infty$.
We see that the arguments of \cite{LT98} go through unchanged:
we shift by half the fundamental domain
$q_4({\vec x}) \rightarrow q_4({\vec x}) - \pi/g\beta =
q^{\prime}_4({\vec
x})$
so that the Jacobian becomes a cosine-squared and the
boundary conditions in $x_4$ of the charged gluons acquire an
extra term: they go from being quasi-periodic to quasi-{\it
anti}periodic.
Next we discretise ${\vec x} = l {\vec n}$,
with directional unit vectors ${\hat e}$, and dimensionless field
$\varphi_{\vec n} \equiv g \beta q^{\prime}_4({\vec x})$.
We obtain for the action
\begin{equation}
S_0^{(l)}[Q] = {l \over {2 g^2 \beta}} \sum_{{\vec n}}
\sum_{{\hat e}}
\varphi_{{\vec n}} [ \varphi_{{\vec n} + 2 {\hat e}}
- 2 \varphi_{{\vec n} + {\hat e}} + \varphi_{\vec n} ]
\; .
\end{equation}
Thus the weight factor appearing in the functional integral is
\begin{equation}
e^{-S^{(l)}_0[Q]} = \sum_{r = 0}^\infty {\cal C}_r ( \frac{l}{g^2
\beta})^r
\ .
\end{equation}
The functional integrals in Eq.(\ref{propdef}) can be done
explicitly
\begin{equation}
\langle 0 | T( q_4({\vec x}) q_4({\vec y}) ) |0 \rangle
 = \frac{1}{4 g^2 \beta^2} ( \frac{\pi^2}{3} - 2)
\delta_{{\vec m}^x, {\vec m}^y}
+ {\cal O}(\frac{l}{g^4 \beta^3}) \; .
\label{qprop}
\end{equation}
So, also in the presence of the self-dual background field,
the correlator of the fluctuating part of Polyakov loops
is ultralocal, being proportional to $\delta^{(3)}({\vec x} -
{\vec y})$
in the continuum limit. The result Eq.(\ref{qprop}) guarantees
that the mass term for the charged gluons is as derived by
\cite{LT98},
namely Eq.(\ref{tempmass}).

We are thus led to an effective action
after integration out of $q_4$ which contains
charged gluon fields $Q_i^{1,2}$ with a mass diverging with
increasing
temperature but, in the presence of the self-dual field,
quasi-antiperiodic boundary conditions.
Moreover, the background field component $B_4(x)$ is still
present
in the action in the usual terms where the original $A_4$
was located, but the field $q_4$ has been successfully integrated
out.

\setcounter{equation}{0}
\renewcommand{\theequation}{C.\arabic{equation}}

\section*{Appendix C: Orthogonality and Completeness Relations.}
In this section we derive the
orthogonality and completeness relations for the
eigenfunctions of the Laplace operator in the presence of the self-dual
homogeneous field and the formulae for the propagator subject to
quasi-antiperiodic boundary conditions. Let us recall first the
solution to
the eigenvalue problem at zero temperature~\cite{Leu81,Eliz}
\begin{eqnarray}
&&-\nabla^2(x)\psi(x)=\lambda\psi(x),
\nonumber\\
&&\nabla^2(x)=(\partial_\mu-iB_\mu(x))^2, \
B_\mu=\frac{1}{2}B_{\mu\nu}x_\nu
\nonumber
\end{eqnarray}
in the space of functions vanishing at infinity.
The operator $-\nabla^2$ can be represented in the form:
\begin{eqnarray}
&&-\nabla^2=2B\left(a^\dagger_\mu Q^-_{\mu\nu}a_\nu+1\right)
\nonumber\\
&&a_\mu=\frac{1}{\sqrt{B}}\left(\frac{1}{2}Bx_\mu+\partial_\mu\right), \
a^\dagger_\mu=\frac{1}{\sqrt{B}}\left(\frac{1}{2}Bx_\mu-\partial_\mu
\right), \
[a_\mu,a^\dagger_\nu]=\delta_{\mu\nu},
\nonumber\\
&&Q^{\pm}_{\mu\nu}=\left(\delta_{\mu\nu}\pm i
b_{\mu\nu}\right)/2, \
Q^\pm Q^\pm=Q^\pm, \ Q^\mp Q^\pm=0, \
b_{\mu\nu}=B_{\mu\nu}/B.
\nonumber
\end{eqnarray}
The matrix $(ib_{\mu\nu})$ can be diagonalised by means of an
appropriate
unitary transformation $U$:
\begin{eqnarray}
&&U^\dagger ib U={\rm diag}(1,-1,1,-1), \ U^\dagger a=\alpha,
 \ [\alpha_\mu,\alpha^\dagger_\nu]=\delta_{\mu\nu},
\nonumber\\
&&\alpha_1=(a_1+ia_2)/\sqrt{2}, \
\alpha_2=(a_1-ia_2)/\sqrt{2}, \
\alpha_3=(a_3+ia_4)/\sqrt{2}, \
\alpha_4=(a_3-ia_4)/\sqrt{2}. \nonumber
\end{eqnarray}
The eigenvalue problem then takes the form:
\begin{eqnarray}
&&2B\left(\alpha_2^\dagger\alpha_2+\alpha_4^\dagger\alpha_4+1
\right)
\psi(x)=\lambda\psi(x),
\nonumber
\end{eqnarray}
with the solution
\begin{eqnarray}
\label{eigf1}
&&\lambda_{k_1k_2}=2B(k_1+k_2+1),
\nonumber\\
&&\psi_{k_1k_2k_3k_4}=
\frac{1}{\sqrt{k_1!k_2!}}(\alpha_2^\dagger)^{k_1}
(\alpha_4^\dagger)^{k_2}\psi_{00k_3k_4}(x),
\nonumber\\
&&\psi_{00k_3k_4}=\frac{B}{2\pi}
\frac{1}{\sqrt{k_3!k_4!}}(\alpha_1^\dagger)^{k_3}
(\alpha_3^\dagger)^{k_4}\exp\left(-\frac{1}{4}Bx^2\right).
\end{eqnarray}
The orthogonality and completeness relations have the following
form:
\begin{eqnarray}
\label{ort1}
&&\int_{-\infty}^\infty d^4x
\psi^\dagger_{k_1k_2k_3k_4}(x)
\psi_{k_1^\prime k_2^\prime k_3^\prime k_4^\prime }(x)=
\delta_{k_1k_1^\prime}
\delta_{k_2k_2^\prime}
\delta_{k_3k_3^\prime}
\delta_{k_4k_4^\prime},\nonumber\\
&&\sum\limits_{k_1k_2k_3k_4}
\psi^\dagger_{k_1k_2k_3k_4}(x)
\psi_{k_1 k_2 k_3 k_4}(y)=\delta(x-y).
\end{eqnarray}
One sees from Eqs.~(\ref{eigf1}) that the spectrum
of $\nabla^2$ is infinitely degenerate which is a consequence of the
homogeneity of the background field.
To proceed further, it is advantageous to introduce the eigenfunctions
$\phi_{k_1k_2}(x,y)$:
\begin{eqnarray}
\label{eigenf2}
\phi_{k_1k_2}(x,y)&=&\frac{B^2}{4\pi^2}
\sum\limits_{k_3k_4}
\sqrt{\frac{(B/2)^{k_3+k_4}}{k_3!k_4!}}(y_1-iy_2)^{k_3}
(y_3-iy_4)^{k_4}\exp\left(-\frac{1}{4}By^2\right)\psi_{k_1k_2k_3k_4}(x)
\nonumber\\
&=&\frac{B^2}{4\pi^2}
\frac{1}{\sqrt{k_1!k_2!}}(\alpha_2^\dagger)^{k_1}
(\alpha_4^\dagger)^{k_2}
\exp\left(-\frac{1}{4}B(x-y)^2
+\frac{i}{2}x_\mu B_{\mu\nu}y_\nu\right),
\nonumber
\end{eqnarray}
Now the degeneracy is parametrised by the continuous variable $y$.
The function $\phi_{00}(x,y)$
can be seen as a matrix element of the projector
onto the subspace spanned by the lowest mode ($k_1=k_2=0$).

Making use of Eqs.~(\ref{ort1}), we arrive at the following
equations for the case of infinite $\beta$ (zero temperature):
\begin{eqnarray}
&&-\nabla^2(x)\Delta(x,y)=\delta(x-y)
\nonumber\\
&&-\nabla^2(x)\phi_{k_1k_2}(x,y)=\lambda_{k_1k_2}\phi_{k_1k_2}
(x,y),
\nonumber\\
&&\phi_{k_1k_2}(x,y)=\phi_{k_1k_2}(x-y)
\exp\{\frac{i}{2}x_\mu B_{\mu\nu}y_\nu\},
\nonumber\\
&&\sum\limits_{k_1k_2} \int_{-\infty}^\infty d^4y
\phi^\dagger_{k_1k_2}(x,y)
\phi_{k_1 k_2}(z,y)=\delta(z-x),
\nonumber\\
&&\int_{-\infty}^\infty d^4x
\phi^\dagger_{k_1k_2}(x,y)
\phi_{k_1^\prime k_2^\prime }(x,z)=
\delta_{k_1k_1^\prime}
\delta_{k_2k_2^\prime}
\phi_{00}(y,z).
\nonumber
\end{eqnarray}
Together with Eq.~(\ref{eigenf2}),
these define respectively the Green's function and eigenfunctions
for the operator $\nabla^2$ as well as completeness and orthogonality
relations for the eigenfunctions.  The propagator can be decomposed
into a sum over projectors onto the subspaces corresponding to
the different eigen-numbers
\begin{eqnarray}
&&\Delta(z,x)=
\sum\limits_{k_1k_2}
\frac{{\cal P}_{k_1 k_2}(z,x)}{\lambda^2_{k_1k_2}},
\nonumber\\
&&{\cal P}_{k_1k_2}(z,x)=
\int_{-\infty}^\infty d^4y
\phi_{k_1 k_2}(z,y)
\phi^\dagger_{k_1k_2}(x,y)
={\cal P}_{k_1k_2}(z-x)e^{\frac{i}{2}z_\mu B_{\mu\nu}x_\nu}.
\end{eqnarray}
The completeness, for instance, is derived in the following way:
\begin{eqnarray}
&&\int_{-\infty}^\infty
d^4y\sum_{k_1k_2}\phi^\dagger_{k_1k_2}(x,y)\phi_{k_1k_2}(z,y)=
\nonumber\\
&&=
\frac{B^2}{4\pi^2}\sum_{k_1k_2k_3k_4}
   \psi^\dagger_{k_1k_2k_3k_4}(x)\psi_{k_1k_2k_3k_4}(z)
\nonumber\\
&&\times\int_{-\infty}^\infty d^4y
\frac{(B/2)^{k_3+k_4}}{k_3!k_4!}
(y_1^2+y_2^2)^{k_3}(y_3^2+y_4^2)^{k_4}e^{-\frac{1}{2}By^2}
\nonumber\\
&&=
\sum_{k_1k_2k_3k_4}
\psi^\dagger_{k_1k_2k_3k_4}(x)\psi_{k_1k_2k_3k_4}(z)=\delta(z-x).
\end{eqnarray}

Then at finite $\beta$ the function
\begin{eqnarray}
\label{db}
D^\beta(x,y)=\sum\limits_{n=-\infty}^{\infty}(-1)^n
\Delta(x_4-y_4+n\beta;\vec x-\vec y)
\exp\left\{\frac{i}{2}x_\mu B_{\mu\nu}y_\nu+in\beta B_4(\vec x+\vec
y)\right\}
\end{eqnarray}
is a solution to the equation
\begin{eqnarray}
&&-\nabla^2(x)D^\beta(x,y)=\delta_\beta(x,y),
\nonumber\\
&&\delta_\beta(x,y)=
\sum\limits_{n=-\infty}^{\infty}(-1)^n
\delta(x_4-y_4+n\beta)\delta(\vec x- \vec y)
\exp\left\{\frac{i}{2}x_\mu B_{\mu\nu}y_\nu+in\beta B_4(\vec x+\vec
y)\right\},
\nonumber
\end{eqnarray}
satisfying the boundary conditions
\begin{eqnarray}
&&D^\beta(x_4+\beta,\vec x;y)=-D^\beta(x_4,\vec x;y)
\exp\left\{-i\beta B_4(\vec x)\right\},
\nonumber\\
&&D^\beta(x;y_4+\beta,\vec y)=-D^\beta(x:y_4,\vec y)
\exp\left\{i\beta B_4(\vec y)\right\}.\nonumber
\end{eqnarray}
Here $\delta_\beta(x,y)$ is the $\delta$-function on the linear
space
$\Phi_\beta$ of functions $f_\beta(x)$ obeying the boundary
condition
$$f_\beta(x_4+\beta,\vec x)=-f_\beta(x_4,\vec x)
\exp\left\{-i\beta B_4(\vec x)\right\}.$$
One can check that completeness is satisfied:
\begin{eqnarray}
\int_{-\infty}^\infty d^3y\int_0^\beta dy_4
\delta_\beta(x,y)f_\beta(y)=f_\beta(x).\nonumber
\end{eqnarray}
Moreover, the functions
\begin{eqnarray}
\phi_{k_1k_2}^\beta(x,y)=
\sum\limits_{n=-\infty}^{\infty}(-1)^n
\phi_{k_1k_2}(x_4-y_4+n\beta,\vec x-\vec y)
\exp\left\{\frac{i}{2}x_\mu B_{\mu\nu}y_\nu+in\beta B_4(\vec x+\vec
y)\right\}
\nonumber
\end{eqnarray}
being the eigenfunctions of $\nabla^2(x)$,
\begin{eqnarray}
-\nabla^2(x)\phi^\beta_{k_1k_2}(x,y)=
\lambda_{k_1k_2}\phi^\beta_{k_1k_2}(x,y),\nonumber
\end{eqnarray}
and satisfying the boundary condition
$$\phi_{k_1k_2}^\beta(x_4+\beta,\vec x;y)=
-\phi_{k_1k_2}^\beta(x_4,\vec x)\exp\left\{-i\beta B_4(\vec
x)\right\},$$
then give
\begin{eqnarray}
&&
\sum\limits_{k_1k_2=0}^{\infty}\int_{-\infty}^{+\infty}
d^3y
\int_0^\beta dy_4\phi^{\beta \dagger}_{k_1k_2}(x,y)
\phi^\beta_{k_1k_2}(z,y)=
\delta_\beta(z,x),
\nonumber\\
&&\int_{-\infty}^{+\infty}d^3x
\int_0^\beta dx_4\phi^{\beta\dagger}_{k_1k_2}(x,y)
\phi^\beta_{l_1l_2}(x,z)=
\delta_{k_1l_1}\delta_{k_2l_2}
\phi^\beta_{00}(y,z)
\nonumber
\end{eqnarray}
so that they define an orthogonal complete basis for the space
$\Phi_\beta$.
The propagator $D^\beta$ can be decomposed over projectors
\begin{eqnarray}
&&D^\beta(z,x)=
\sum\limits_{k_1k_2}
\frac{{\cal P}^\beta_{k_1 k_2}(z,x)}{\lambda^2_{k_1k_2}},
\nonumber\\
&&P^\beta_{k_1k_2}(z,x)=
\int\limits_{-\infty}^\infty d^3y
\int\limits_{0}^\beta dy_4
\phi_{k_1 k_2}(z,y)
\phi^\dagger_{k_1k_2}(x,y)
\nonumber\\
&&=\sum\limits_{n=-\infty}^\infty (-1)^n
{\cal P}_{k_1k_2}(z_4-x_4+n\beta, \vec z - \vec x)
\exp\left\{\frac{i}{2}z_\mu B_{\mu\nu}x_\nu+im\beta B_4(\vec z+\vec x)
\right\}.
\end{eqnarray}
Taking into account the representation of the zero temperature
propagator in terms of the projector operators
we get Eq.~(\ref{db}).

Let us show how the completeness of the set can be derived.
We have to evaluate the integral
\begin{eqnarray}
&&
\sum\limits_{k_1k_2=0}^{\infty}\int_{-\infty}^{+\infty}
d^3y
\int_0^\beta dy_4\phi^{\beta \dagger}_{k_1k_2}(x,y)
\phi^\beta_{k_1k_2}(z,y)=
\sum\limits_{k_1k_2=0}^{\infty}\int_{-\infty}^{+\infty}
d^3y\int_0^\beta dy_4
\sum\limits_{n,m=-\infty}^{\infty}(-1)^{n+m}
\nonumber\\
&&
\times\phi^{\dagger}_{k_1k_2}(x_4-y_4+n\beta,\vec x-\vec y)
\phi_{k_1k_2}(z_4-y_4+m\beta,\vec z -\vec y)
\nonumber\\
&&\times\exp\left\{
-\frac{i}{2}x_\mu B_{\mu\nu}y_\nu
+\frac{i}{2}z_\mu B_{\mu\nu}y_\nu
-in\beta B_4(\vec x+\vec y)+im\beta B_4(\vec z+\vec y)
\right\}.
\nonumber
\end{eqnarray}
After the change of integration variable
$ y^\prime_4=y_4-n\beta$ we get for the right hand side of this equation:
\begin{eqnarray}
&&\sum\limits_{k_1k_2=0}^{\infty}\int_{-\infty}^{+\infty}d^3y
\sum\limits_{n,m=-\infty}^{\infty}(-1)^{n+m}
\int_{-n\beta}^{(1-n)\beta} dy^\prime_4
\nonumber\\
&&
\times\phi^{\dagger}_{k_1k_2}(x_4-y^\prime_4,\vec x-\vec y)
\phi_{k_1k_2}(z_4-y^\prime_4+(m-n)\beta,\vec z -\vec y)
\nonumber\\
&&\times\exp\left\{
-\frac{i}{2}x_\mu B_{\mu\nu}y^\prime_\nu
+\frac{i}{2}z_\mu B_{\mu\nu}y^\prime_\nu
+i(m-n)\beta B_4(\vec z+\vec y)
\right\}.
\nonumber
\end{eqnarray}
Finally, shifting the variable $m^\prime=m-n$ in the sum and denoting
$z_4^\prime=z_4+m^\prime\beta$ we arrive at
\begin{eqnarray}
&&
\sum\limits_{k_1k_2=0}^{\infty}\int_{-\infty}^{+\infty}
d^3y
\int_0^\beta dy_4\phi^{\beta \dagger}_{k_1k_2}(x,y)
\phi^\beta_{k_1k_2}(z,y)
\nonumber\\
&&=\sum\limits_{m^\prime=-\infty}^{\infty}(-1)^{m^\prime}
\sum\limits_{k_1k_2=0}^{\infty}\int_{-\infty}^{+\infty}d^3y
\int_{-\infty}^{\infty} dy_4
\phi^{\dagger}_{k_1k_2}(x,y)
\phi_{k_1k_2}(z^\prime,y)
\exp\left\{
im^\prime\beta B_4(\vec z)
\right\}
\nonumber\\
&&=
\sum\limits_{m=-\infty}^{\infty}(-1)^m
\delta(z_4-x_4+m\beta)\delta(\vec z- \vec x)
\exp\left\{\frac{i}{2}z_\mu B_{\mu\nu}x_\nu+im\beta B_4(\vec z+\vec
x)\right\}=
\delta_\beta(z,x).
\nonumber
\end{eqnarray}
The orthogonality and decomposition of the propagator over the projectors
can be obtained in a similar way.

\setcounter{equation}{0}
\renewcommand{\theequation}{D.\arabic{equation}}

\section*{Appendix D: Gluon Propagator in Self-Dual Background
Field in the Temporal Gauge}
We start with the function
\begin{eqnarray}
D_{ij}(x,y|B,m)=
\Delta_{ij}(x_4-y_4;{\vec x} -{\vec y} |B,m)
\exp\left(\frac{i}{2}x_\mu B_{\mu\nu}y_\nu\right)
\nonumber
\end{eqnarray}
which is a solution of the equation
\begin{eqnarray}
\left[\left(-\nabla^2+M^2\right)\delta_{ij}+D_i D_j
-2iB_{ij}\right]
D_{jk}(x,y|B,m)=\delta_{ik}\delta(x-y)
\nonumber
\end{eqnarray}
in the limit of infinite $\beta$.
The function $\Delta$ is then a solution to the equation
\begin{eqnarray}
%\label{green2}
\left[\left(-\nabla^2+M^2\right)\delta_{ij}+D_i D_j
-2iB_{ij}\right]
\Delta_{jk}(x|B,m)=\delta_{ik}\delta(x). \nonumber
\end{eqnarray}
The matrix $iB_{ij}$ can be diagonalised by an appropriate
unitary transformation $U$, so that we arrive at
\begin{eqnarray}
 \label{green3}
\left[\left(-\nabla^2+m^2\right)\delta_{rs}+\tilde D_r \tilde
D_s^\prime
-2B\delta_{rs}\xi_s\right]
\tilde\Delta_{st}(x|B,m)=\delta_{rt}\delta(x)
\end{eqnarray}
with $r,s,t \in \{ 0,1,-1 \}$ and
$\xi_s=s$ the gluon spin projections onto the third spatial axis.
Moreover,
$$\nabla^2=D_4^2+\tilde D_r^\prime\tilde D_r, \ \
\tilde D_s=U^\dagger_{sj}D_j, \ \
\tilde D_s^\prime=D_j U_{js}.
$$

We next decompose the propagator as
\begin{eqnarray}
\label{general}
\tilde \Delta_{rs}=\delta_{rs}F_s +
\tilde D_r \tilde D_s^\prime H_s +
i\delta_{r0}\tilde D_s^\prime L + i\delta_{s0} \tilde D_r N +
\delta_{r0}\delta_{s0}P.
\end{eqnarray}
Using this and the relations
\begin{eqnarray}
&& [D_4^2,\tilde D_s]=2iB\delta_{s0}D_4, \ \
[\tilde D_r,\tilde D_s^\prime\tilde D_s]=2B\delta_{rt}\xi_t\tilde
D_t,
\nonumber\\
&& [D_4,\tilde D_s]=[D_4,\tilde D^\prime_s]=iB\delta_{s0}, \ \
[\tilde D_0,\tilde D_s]=0, \nonumber\\
&& \sum_j\left[-\tilde D^2\delta_{rs}+\tilde D_r \tilde
D_s^\prime
-2B\delta_{rs}\xi_s\right]\tilde D_s\equiv 0,
\nonumber
\end{eqnarray}
we can rewrite Eq.~(\ref{green3}) as a system of differential
equations
for the functions $F$, $H$, $L$, $N$, $P$:
\begin{eqnarray}
\label{system1}
&&\left(-\nabla^2-2B\xi_s+m^2\right)F_s(x)=\delta(x),
\nonumber\\
&&i\tilde D_0L_s(x)+\left(-D_4^2+m^2\right)H_s(x)+F_s(x)=0,
\nonumber\\
&&-2iB D_4H_s(x)+\left(-\nabla^2-2B\xi_s+m^2\right)L_s(x)=0,
\nonumber\\
&&-i\tilde D_0^\prime P_s(x)-2iB
D_4H_s(x)-\left(-D_4^2+m^2\right)N_s(x)=0,
\nonumber\\
&&2iB D_4\left(L_s(x)-N_s(x)\right)+2B^2H_s(x)
+\left(-\nabla^2+m^2\right)P_s(x)=0.
\end{eqnarray}
We now show that, for the calculation of the space-time trace of
the propagator, we only need to know the functions $H$, $L$, $N$
and $P$
in the neighborhood of $x=0$.
Consider the integral
\begin{eqnarray}
&&{\cal T}_{jk}=\int_{V} d^3x\int_0^{\beta}dx_4
D^{\beta}_{jk}(x,x|B,m),
\nonumber
\end{eqnarray}
which is contained in Eq.~(\ref{p-eff}). With $L_3$ the
length of the third space direction and using
Eqs.~(\ref{l-prop},\ref{str})
we get
\begin{eqnarray}
\label{surprise}
&&{\cal T}_{jk}=\int_{V} d^3x\int_0^{\beta}dx_4
\sum\limits_{n=-\infty}^{\infty}(-1)^n
\Delta_{jk}(n\beta;0|B,m)
\exp\left(\mp \frac{i}{2}n\beta B x_3\right)
\nonumber\\
&&=V \beta\left[
\Delta_{jk}(0;0|B,m) + 4\lim_{L_3\to\infty}
\sum\limits_{n=1}^{\infty}(-1)^n
\Delta_{jk}(n\beta;0|B,m)
\frac{\sin\left(\frac{1}{2}nBL_3 \beta \right)}{nBL_3\beta }\right]
\nonumber\\
&&=V \beta\left[
\Delta_{jk}(0;0|B,m) + O\left(L^{-1}_3 \beta^{-1}\right)
\right].
\end{eqnarray}
Equation (\ref{surprise}) leads to the result that the
terms with $n\not=0$ do not contribute to the effective
potential.
Further calculations can be simplified due to this
property of the external field Eq.(\ref{field}).
The solution of Eqs.~(\ref{system1}) depends on $x^2$, and
the first order derivatives are proportional to
$x_0=U^\dagger_{0j}x_j$ or $x_4$.
If we need the propagator only for $x\to0$, we can omit all terms
which contain the first order derivatives.
Thus, in the limit $x\to0$, we have to solve the equations
\begin{eqnarray}
\label{system2}
&&\left(-\nabla^2-2B\xi_s+m^2\right)F_s(x)=\delta(x),
\nonumber\\
&&\left(-D_4^2+m^2\right)H_s(x)+F_s(x)=0,
\nonumber\\
&&\left(-\nabla^2-2B\xi_s+m^2\right)L_s(x)=0,
\nonumber\\
&&\left(-D_4^2+m^2\right)N_s(x)=0,
\nonumber\\
&&\left(-\nabla^2+m^2\right)P_s(x)+2B^2H_s(x)=0. \nonumber
\end{eqnarray}
One can check that the positive-definiteness of the spectrum of
the
operators
$(-D_4^2+m^2)$ and $(-\nabla^2-2B\xi_s+m^2)$
in the space of functions vanishing at infinity.
This means that $L_s(x)\to0$ and $N_s(x)\to0$ for $x\to 0$.
Finally one gets for small $x^2$
\begin{eqnarray}
\label{system3}
&&F_s(x)=\left(-\nabla^2-2B\xi_s+m^2\right)^{-1}\delta(x),
\nonumber\\
&&H_s(x)=-\left(-D_4^2+m^2\right)^{-1}F_s(x),
\nonumber\\
&&P_s(x)=-2B^2\left(-\nabla^2+m^2\right)^{-1}H_s(x),
\nonumber\\
&&L_s(x)=0,
\ \ N_s(x)=0.
\end{eqnarray}
Using Eqs.~(\ref{system3}) and (\ref{general}), we arrive at the
relation
Eq.(\ref{tr1}) given in the main body of the paper.
We have only to insert the delta-function into the equations
for $H_s$ and $P_s$ and to represent these functions as
convolutions
of the propagators $F_k$, $F_0$ and $\Delta_4$, where the latter
is the Green's function corresponding to $(-D_4^2 + m^2)$.
These lead to the expressions in Eq.(\ref{fk}).

\end{document}